\documentclass[traditabstract]{aa}
\usepackage{graphicx}

\usepackage{epsfig}
\usepackage[colorlinks=true,linkcolor=blue]{hyperref}

\newcommand{\be}{\begin{equation}}
\newcommand{\ee}{\end{equation}}

\newcommand{\bea}{\begin{eqnarray}}
\newcommand{\eea}{\end{eqnarray}}

\newcommand{\al}{\alpha}
\newcommand{\bt}{\beta}

\newcommand{\Gm}{\Gamma}
\newcommand{\dl}{\delta}
\newcommand{\Dl}{\Delta}

\newcommand{\et}{\eta}

\newcommand{\lm}{\lambda}
\newcommand{\Lm}{\Lambda}

\newcommand{\rh}{\rho}

\newcommand{\sg}{\sigma}

\newcommand{\ups}{\upsilon}

\newcommand{\Om}{\Omega}

\newcommand{\rarrow}{\rightarrow}

\newcommand{\nn}{\nonumber}

\newcommand{\varep}{\varepsilon}

\begin{document}

\title{Cosmological perturbations in the $\Lm$CDM-like limit of a polytropic dark matter model}

\author{K. Kleidis \inst{1} \and N. K. Spyrou \inst{2}}

\institute{Department of Mechanical Engineering, Technological Education Institute of Central Macedonia, \\621.24 Serres, Greece. \\e-mail: \texttt{kleidis@teiser.gr} \and Department of Astronomy, Aristoteleion University of Thessaloniki, 541.24 Thessaloniki, Greece. \\e-mail: \texttt{spyrou@auth.gr}}

\offprints{N. K. Spyrou}

\date{Received ..... ; accepted .....}

\titlerunning{Polytropic DM perturbations}

\authorrunning{K. Kleidis \& N. K. Spyrou}

\abstract{In a recent article, Kleidis \& Spyrou (2015) proposed that both dark matter (DM) and dark energy (DE) can be treated as a single component, if accommodated in the context of a polytropic DM fluid with thermodynamical content. Depending on only one free parameter, the polytropic exponent, $-0.103 < \Gm \leq 0$, this unified DM model reproduces to high accuracy the distance measurements performed with the aid of the supernovae Type Ia (SNe Ia) standard candles, without suffering (either from the age problem or) from the coincidence problem, i.e., interpreting not only when but also why the Universe transits from deceleration to acceleration so recently. Still, however, there is a critical issue that the polytropic DM model should also confront with, that is, to demonstrate its compatibility with current observational data concerning structure formation. To begin unfolding (also) this knot, in the present article we discuss the evolution of cosmological perturbations in the $\Lm$CDM-like (i.e., $\Gm = 0$) limit of the polytropic DM model. The corresponding results are quite encouraging, since, such a model reproduces every major effect already known from conventional (i.e., pressureless cold dark matter - CDM) structure formation theory, such as the constancy of metric perturbations in the vicinity of recombination and the (late-time) Meszaros effect on their rest-mass density counterparts (Meszaros 1974). The non-zero (polytropic) pressure, on the other hand, drives the evolution of small-scale velocity perturbations along the lines of the root-mean-square velocity law of conventional Statistical Physics. As a consequence, in this model "peculiar velocities" slightly increase, instead of being redshifted away by cosmic expansion, a result that might comprise a convenient probe of the polytropic DM model with $\Gm = 0$. What is more important is that, upon consideration of scale-invariant metric perturbations, the spectrum of their rest-mass density counterparts exhibits an effective power-law dependence on the (physical) wavenumber, $k_{ph}$, of the form $k_{ph}^{3+n_s^{eff}}$, with the associated scalar spectral index, $n_s^{eff}$, being equal to $n_s^{eff} = 0.970$; a theoretical value that actually reproduces the corresponding observational (Planck) result, i.e., $n_s^{obs} = 0.968 \pm 0.006$ (Ade et al. 2016).}

\keywords{cosmological parameters - dark energy - dark matter - cosmology: theory}

\maketitle

\section{Introduction}

Currently, an extended list of observational data suggests that, apart from the DM abundance and a (small) baryon contamination, the Universe contains also a uniformly distributed energy component with negative pressure, which, at relatively low values of cosmological redshift, $z$, drives the cosmic expansion, causing its acceleration (see, e.g., Olive et al. 2014, p. 361). Reflecting our ignorance on its exact nature, this new component - which constitutes about two-thirds of the Universe mass-energy content - was termed dark energy - DE (Turner \& White 1997; Perlmutter et al. 1999$b$). 

The need for DE was first suggested by high-precision distance measurements, performed with the aid of the SNe Ia standard candles (Hamuy et al. 1996; Garnavich et al. 1998; Perlmutter et al. 1998, 1999$a$; Schmidt et al. 1998; Riess et al. 1998, 2001, 2004, 2007;  Knop et al. 2003; Tonry et al. 2003; Barris et al. 2004; Krisciunas et al. 2005; Astier et al. 2006; Jha et al. 2006; Miknaitis et al. 2007; Wood-Vasey et al. 2007; Amanullah et al. 2008, 2010; Holtzman et al. 2008; Kowalski et al. 2008; Hicken et al. 2009$a$, 2009$b$; Kessler et al. 2009; Contreras et al. 2010; Guy et al. 2010; Suzuki et al. 2012). Today, there is also evidence from galaxy clusters (Allen et al. 2004), the integrated Sachs-Wolfe (ISW) effect (Boughn \& Crittenden 2004), baryon acoustic oscillations (BAOs, Eisenstein et al. 2005; Percival et al. 2010), weak gravitational lensing (WGL, Huterer 2002; Copeland et al. 2006), and the Lyman-$\al$ (LYA) forest (Seljak et al. 2006). A combination of these data with those from the Wilkinson microwave anisotropy probe (WMAP) survey (see, e.g., Komatsu et al. 2011; Bennett et al. 2013) has provided evidence for cosmic acceleration (and, hence, for DE, as well) at the $5 \sg$ confidence level.

Although the notion of DE can be attributed to a non-vanishing cosmological constant, $\Lm$ (see, e.g., Riess et al. 1998; Perlmutter et al. 1999$a$), such a choice fails to explain the magnitude of $\Lm$ itself, since the corresponding theoretical prediction are $10^{123}$ times larger than what is observed (cf. Sahni \& Starobinsky 2000; Padmanabhan 2003). As a consequence, many other physically-motivated models have appeared in the literature, including quintessence (Caldwell et al. 1998), k-essence (Armendariz-Picon et al. 2001), phantom cosmology (Caldwell 2002) and tachyonic matter (Padmanabhan 2002), involving (also) several braneworld scenarios, such as DGP-gravity (Dvali et al. 2000) and the landscape scenario (Bousso \& Polchinski 2000), as well as alternative-gravity theories, such as the scalar-tensor theories (Esposito-Farese \& Polarski 2001), $f(R)$-gravity (Capozziello et al. 2003) and modified gravity (Nojiri \& Odintsov 2007), holographic gravity (Cohen et al. 1999; Li 2004; Pav\'{o}n \& Zimdahl 2005), Chaplygin gas (Kamenshchik et al. 2001; Bento et al. 2002; Bean \& Dor\'{e} 2003; Sen \& Scherrer 2005), Cardassian cosmology (Freese \& Lewis 2002; Gondolo \& Freese 2003; Wang et al. 2003), theories of compactified internal dimensions (Mongan 2001; Defayet et al. 2002; Perivolaropoulos 2003; Sami et al. 2004), mass-varying neutrinos (Fardon et al. 2004; Peccei 2005), and so on (for a detailed review on the various DE models see, e.g., Caldwell \& Kamionkowski 2009; Miao et al. 2011).

Among them, the so-called unified DM models (see, e.g., Zimdahl et al. 2001; Bili\'{c} et al. 2002; Balakin et al. 2003; Gondolo \& Freese 2003; Makler et al. 2003; Scherrer 2004; Ren \& Meng 2006; Meng et al. 2007; Lima et al. 2008, 2010, 2012; Basilakos \& Plionis 2009, 2010; Dutta \& Scherrer 2010; Xu et al. 2012) have attracted much of attention. In view of these models, both DM and DE can be described in terms of a single component, namely, the self-interacting DM (see, e.g., Spergel \& Steinhardt 2000; Arkani-Hamed et al. 2009; Cirelli et al. 2009; Cohen \& Zurek 2010; Van den Aarssen et al. 2012). In fact, such a possibility is strongly suggested by observational data from the WMAP survey (Hooper et al. 2007), being further reenforced by recent results from the high-energy particle detector PAMELA (Adriani et al. 2009, 2010), the Alpha Magnetic Spectrometer (AMS) on the International Space Station (Aguilar et al. 2016) and the Fermi Large Area Telescope (LAT) survey (Albert et al. 2016).

In this context, in a recent work by Kleidis \& Spyrou (2015) it was proposed that the self-interacting DM could (at least phenomenologically) attribute to the Universe matter content some sort of fluid-like properties, and consequently lead to a conventional approach to the DE concept. Indeed, if the DM constituents collided with each other frequently enough, thus enabling their (kinetic) energy to be redistributed, a uniform extra-energy component might be present in the Universe, given by the energy of the internal motions of a thermodynamically involved DM fluid. As to what sort of thermodynamical processes could lead to a description that would be compatible to the observed characteristics of the accelerating Universe, the answer is, those of the polytropic kind.

Polytropic processes in a DM fluid have been most successfully used in modeling dark galactic haloes, improving significantly the velocity dispersion profiles of galaxies (Bharadwaj \& Kar 2003; Nunez et al. 2006; Zavala et al. 2006; B{\"o}hmer \& Harko 2007; Saxton \& Wu 2008; Su \& Chen 2009; Saxton \& Ferreras 2010; Saxton et al. 2016). On cosmological level, polytropic (DM) models were first encountered as natural candidates for Cardassian Cosmologies (see, e.g., Freese \& Lewis 2002; Gondolo \& Freese 2003; Wang et al. 2003; Freese 2005), and they have been widely used as phenomenological models of an extra, self-contained, DE cosmological fluid (see, e.g., Nojiri et al. 2005; Stefanci\'{c} 2005; Mukhopadhyay et al. 2008; Karami et al. 2009; Karami \& Abdolmaleki 2010$a$, 2010$b$, 2012; Malekjani et al. 2011; Chavanis 2012$a$, 2012$b$, 2012$c$; Karami \& Khaledian 2012; Asadzadeh et al. 2013). More recently, polytropic characteristics were also attributed to an effective cosmic fluid obtained in the context of generalized Galileon cosmology (Koutsoumbas et al. 2017).

Our approach, however, is completely different, since it does not involve any extra DE at all. Instead, we have simply examined the dynamical properties of a cosmological model driven by a gravitating (DM) fluid with thermodynamical content, the volume elements of which perform polytropic flows. In this case, the energy of this fluid's internal motions is also taken into account as a source of the universal gravitational field, as it should. As we proved (see Kleidis \& Spyrou 2015, 2016), this form of energy can compensate for the extra energy needed to compromise spatial flatness, namely, to justify that, today, the total energy density parameter is exactly unity. The polytropic (DM) model we proposed, depends on only one free parameter, the polytropic exponent, $-0.103 < \Gm \leq 0$, and does not suffer either from the age problem or from the coincidence problem. At the same time, this model reproduces to high accuracy the distance measurements performed with the aid of the SNe Ia standard candles, without the need for any exotic DE or the cosmological constant. Finally, the polytropic DM model most natutally interprets, not only when, but also, why the Universe transits from deceleration to acceleration so recently, thus rising as a mighty contestant for a realistic (though conventional) DE model. Still, however, there is a critical issue that this model should also confront with, that is to demonstrate its compatibility with current observational data concerning structure formation. 

The theory of structure formation is based on gravitational instability and aims at describing how primordial fluctuations in matter grow into galaxies and cluster of galaxies, due to self-gravity. A perturbative approach can be used when the amplitudes of these fluctuations are small; hence, their growth can be solved to linear approximation. Cosmological perturbations over a homogeneous and isotropic background, as they historically developed (Lifshitz 1946; Lifshitz \& Khalatnikov 1963; Hawking 1966; Sachs \& Wolfe 1967; Weinberg 1972; Peebles 1980; Mukhanov et al. 1992; Padmanabhan 1993), are well-suited to describe a curved background that is filled with ordinary matter, the stress-energy tensor of which can be described in terms of an equation of state. Such a theory, has been most successfully used to study the growth of inhomogeneities in radiation, baryonic matter, and DM (see, e.g., Ma \& Bertschinger 1995). 

At this point, we should note that, as far as structure formation is concerned, all forms of DM are not equivalent. Particles that are highly relativistic (such as neutrinos or other particles with masses lower than $100 \: eV/c^2$) have the property that, due to free streaming, erase perturbations out to very large scales (Bond et al. 1980). In this case, very-large-scale structures form first and subsequently fragment to form galaxies later. Particles with this property are termed hot dark matter (HDM). On the other hand, CDM (i.e., particles with masses larger than $1 \: MeV/c^2$) has the opposite behavior: Small-scale structures form first, aggregating to form larger structures later (Bond \& Szalay 1983). It is now well-known that pure HDM cosmologies can not reproduce the observed large-scale structure of the Universe (see, e.g., Klypin et al. 1993), in contrast to the CDM ones, which, however, are considered to be pressureless. As a consequence, the evolution of cosmological perturbations in CDM models has been restricted in the Newtonian regime (see, e.g., Veeraraghavan \& Stebbins 1990; Knobel 2012, pp. 74, 75). 

Nevertheless, in the polytropic DM approach, we do not neglect the pressure, $p$, with respect to the overall-energy density, $\varep$, hence, even well-within the matter-dominated era, cosmological perturbations should be treated in a general-relativistic manner. In this context, in order to define the energy density perturbation, one needs to choose a particular set of hypersurfaces, i.e., one needs to choose a gauge. The (so-called) Newtonian gauge (Mukhanov et al. 1992) is a particularly convenient choice to treat scalar perturbations, since, in this case, the scalar fields that describe the metric perturbations are identical (up to a minus sign) to the gauge-invariant variables introduced by Bardeen (1980). For this reason, in the present article we adopt the Newtonian gauge. 

This paper is organized as follows. In Sect. 2, we summarize the basic features of the polytropic DM model and its $\Lm$CDM-like $(\Gm = 0)$ limit. In Sect. 3 we present the mathematical setup that leads to the system of differential equations which govern the evolution of small-scale fluctuations in a (dark-) matter-dominated Universe, perturbed over a spatially-flat Friedmann-Robertson-Walker (FRW) model, the (zeroth order) evolution of which is driven by a polytropic (DM) fluid with thermodynamical content. In Sect. 4, we restrict ourselves to the $\Lm$CDM-like (i.e., $\Gm = 0$) limit of this model. As we find, in this case, the perturbations' equations decouple, and can be solved analytically, to give the form of the generalized Newtonian potential, $\phi$, as a function of the cosmological redshift, $z$. Accordingly, the rest-mass density contrast, $\dl$, and the comoving counterpart, $\ups$, of the "peculiar" velocity field, $\ups_{pec}$, are also obtained as functions of $z$. A subsequent analysis of these results, shows that, our solution for $\lbrace \phi, \dl \rbrace$ can reproduce every major effect already known from conventional (i.e., pressureless CDM) cosmological perturbations' theory. In particular, at $100 \leq z \leq 1090$, the generalized Newtonian potential is $\vert \phi \vert \approx constant$, justifying the current scientific perception that, during the early matter-dominated era the metric perturbations were (more or less) constant (see, e.g., Knobel, p. 75). As far as matter perturbations are concerned, the corresponding small-scale modes (i.e., those lying well-within the horizon) conform with the (so-called) Meszaros effect (Meszaros 1974). On the contrary, modes of linear dimensions comparable to horizon's length are suppressed as $\left ( 1 + z \right )^{-1}$, which means that, at relatively large values of $z$ $(100 \leq z \leq 1090)$, only the small-scale structures we see today were allowed to be formed. We expect that, on the approach to the present epoch (i.e., at lower cosmological redshift values), those small-scale strucrures must have subsequently aggregated to form macrostructures, in compatibility to the CDM approach. On the other hand, because of the non-zero polytropic pressure, "peculiar" velocities are no longer redshifted away, as it is predicted by conventional (i.e., pressureless) structure formation theory (see, e.g., Peacock 1999, p. 470; Sparke \& Ghallagher 2007, p. 350). Instead, along the lines of the root-mean-square (rms) velocity law of Statistical Physics, $\ups$ increases in proportion to the square root of the Universe scale factor, $R$, although still remains conveniently small (e.g., for $z \geq 100$, $\ups \leq 10^{-3} \ll 1$). Eventually, in Sect. 5, we explore the dimensionless power spectrum of rest-mass density perturbations in the $\Lm$CDM-like (i.e., $p = constant \neq 0$) limit of the unified DM model under consideration. As we find, provided that the spectrum of metric perturbations is scale invariant, as implied by the cosmic microwave background (CMB) anisotropy measurements (see, e.g., Komatsu et al. 2009, 2011) and several other physical arguments (see, e.g., Padmanabhan 1993, p. 229; Peacock 1999, p. 499), its rest-mass density counterpart exhibits an effective power-law dependence on the (physical) wavenumber, $k_{ph}$, of the form $k_{ph}^{3+n_s^{eff}}$, with the associated scalar spectral index, $n_s^{eff}$, being equal to $n_s^{eff} = 0.970$. It is worth noting that, this value only slightly differs from the corresponding observational (Planck) result, i.e., $n_s^{obs} = 0.968 \pm 0.006$ (Ade et al. 2016). To the best of our knowledge, this is the first time that a conventional model with practically zero free parameters predicts a theoretical result so close to observation. Finally, we conclude in Sect. 6. In what follows, we consider $c = 1 = \hbar$.

\section{Polytropic (DM) semantics}

A few years ago, in the context of a unified DM scenario, Kleidis \& Spyrou (2015) explored the properties and the associated phenomenology of a (spatially-flat) cosmological model in which the fundamental units of the Universe matter-energy content are the volume elements of a collisional DM fluid performing polytropic flows. In this case, the (isotropic) pressure, $p$, of the cosmic fluid is related to its rest-mass density, $\rh$, through the barotropic equation of state (EoS) \be p = p_0 \left ( \frac{\rh}{\rh_0} \right )^{\Gm} \: , \ee where $p_0$ and $\rh_0$ are the associated present-time values and $\Gm$ is the polytropic exponent (in connection, see, e.g., Chandrasekhar 1939, pp. 85-86; Horedt 2004, pp. 5-9).

In terms of this approach, along with all the other physical characteristics, the internal thermodynamic energy of the cosmic fluid should be taken (also) into account as a source of the universal gravitational field. In other words, in a polytropic cosmological model the overall-energy density, $\varep$, of the Universe matter-energy content is no longer given solely by its rest-mass counterpart, $\rh$, but includes (also) an extra term, $\rh {\cal U}$, associated with the energy of the cosmic fluid's internal motions, ${\cal U}$, per unit of specific volume, $\frac{1}{\rh}$ (for a detailed analysis, see, e.g., Fock 1959 pp. 81-83 and 91-94). This form of energy may actually serve as the DE needed to compromise spatial flatness [cf. Kleidis \& Spyrou 2015, Eqs. (49) - (50)], i.e., to justify that, today, the overall-energy density parameter, $\Om$, is very close to unity (see, e.g., Komatsu 2009, 2011), that is much larger than the measured value of its rest-mass counterpart, $\Om_M = 0.308$ (Ade et al. 2016).

Restricting ourselves to a perfect-fluid source, the combination of the continuity equation, \be {\cal T}^{0 \nu}_{; \nu} = 0 \: , \ee where ${\cal T}^{\mu \nu}$ is the energy-momentum tensor of the Universe matter-energy content (Greek indices, $\mu, \: \nu = 0, 1, 2, 3$, refer to four-dimensional spacetime and the semicolon denotes covariant derivative), with the first law of Thermodynamics in curved spacetime, \be d {\cal U} + p d \left ( \frac{1}{\rh} \right ) = {\cal C} dT \: , \ee where $T$ is the polytropic-DM temperature and ${\cal C}$ is the associated specific heat, result in the decomposition of $\varep$ as follows \be \varep = \rh + \frac{1}{\Gm - 1} p \ee [cf. Eq. (39) of Kleidis \& Spyrou 2015], where we have taken (also) into account that, in a closed thermodynamical system (e.g., any volume element of the fluid-like model so considered) the total number of particles is conserved; hence, \be \rh R^3 = constant \: , \ee with $R$ being the Universe scale factor. Accordingly, inserting Eq. (4) into Friedmann equation, \be H^2 = \frac{8 \pi G}{3} \varep \: , \ee where $H = \frac{\dot{R}}{R}$ is the Hubble parameter, $G$ is Newton's universal constant of gravitation and the dot denotes differentiation with respect to cosmic time, $t$, we obtain \be \left ( \frac{H}{H_0} \right )^2 = \Om_M \left ( \frac{R_0}{R} \right )^3 \left [ 1 + \frac{1}{\Gm - 1} \frac{p_0}{\rh_0} \left ( \frac{R_0}{R} \right )^{3 (\Gm - 1)} \right ] \ee [cf. Eq. (43) of Kleidis \& Spyrou 2015], with $H_0$ and $R_0$ representing the present-time values of $H$ and $R$, respectively. As a consequence, today, Eq. (7) is reduced to \be p_0 = \rh_0 (\Gm - 1) \frac{1 - \Om_M}{\Om_M} \: . \ee In view of Eq. (8), for $\Gm < 1$, the pressure given by Eq. (1) becomes negative, and so does the quantity $\varep + 3 p$ at cosmological redshifts, $z \equiv \frac{R_0}{R} - 1$, lower than a transition value, $z_{tr}$ [cf. Eqs. (107) - (108) of Kleidis \& Spyrou 2015]. In other words, for $z \leq z_{tr}$, a cosmological model filled with polytropic DM fluid accelerates its expansion. Indeed, upon consideration of Eq. (8), the Friedmann equation (7) reads \be \left ( \frac{H}{H_0} \right )^2 = \Om_M \left ( 1 + z \right )^3 + (1 - \Om_M) \left ( 1 + z \right )^{3 \Gm} \: , \ee representing a (spatially-flat) cosmological model filled with CDM and  (dynamically evolving) DE [see, e.g., Linder \& Jenkins 2003, Eq. (2); Amendola et al. 2013, Eq. (1.3.1)], the amount of which, at the present epoch, rises to $1 - \Om_M = 0.692$ of the Universe matter-energy content (Ade et al. 2016). 

It is worth noting that, for $\Gm = 0$, the theoretically derived value for $z_{tr}$ [cf. Eq. (72) of Kleidis \& Spyrou 2015] actually reproduces the corresponding $\Lm$CDM result [see, e.g., Capozziello et al. 2015, Eq. (20)]. In fact, for $\Gm = 0$, the polytropic cosmological model under consideration reproduces every major aspect of the widely admitted $\Lm$CDM model [cf. Eq. (9), above], such as the Universe expansion law, $R (t) \sim \sinh^{2/3} t$ (see, e.g., Frieman et al. 2008, p. 6), and the associated "age", $t_0 \approx 13.80$ billion years [cf. Kleidis \& Spyrou 2015, Eqs. (64) and (66), respectively]. For $\Om_M = 0.308$ (Ade et al. 2016), it (furthermore) reproduces the present-time value, $q_0$, of the deceleration parameter, $q$, predicting $q_0 = - 0.54$, whereas current observational data suggest that $q_0 = - 0.53_{-0.13}^{+0.17}$ (see, e.g., Giostri et al. 2012), and the $\Lm$CDM-oriented value of the CMB-shift parameter, in which case, the $\Gm = 0$ limit of the polytropic DM model predicts ${\cal R} = 1.7342$ [cf. Kleidis \& Spyrou 2015, Eq. (112)], whereas the nine-year WMAP survey (Bennett et al. 2013) suggests that ${\cal R} = 1.7329 \pm 0.0058 \, (68 \% \, CL)$. In view of all the above, in what follows, the polytropic DM model with $\Gm = 0$ will be termed as "$\Lm$CDM-like limit of the polytropic DM model".

In this limit, the polytropic DM model under consideration fully compromises the parameterization of the (so-called) total EoS parameter, \be w_{tot} \equiv \frac{p}{\varep} \: , \ee in terms of $z$. Indeed, for $\Gm = 0$, upon consideration of Eqs. (4), (5) and (8), Eq. (10) yields \be w_{tot} \equiv \frac{p}{\varep} = - \frac{1 - \Om_M}{1 - \Om_M + \Om_M (1 + z)^3} \: , \ee the evolution of which, as a function of $z$, is presented in Fig. 1. We observe that, today, i.e., for $z = 0$, $w_{tot} = - \left ( 1 - \Om_M \right ) = - 0.692$, in complete agreement to the corresponding $\Lm$CDM result, \bea w_{tot} = \frac{p_{tot}}{\rh_{tot}} & = & \frac{p_{\Lm}}{\rh_M + \rh_{\Lm}} = \frac{- \rh_{\Lm}}{\rh_M + \rh_{\Lm}} = \frac{- \Om_{\Lm}}{\Om_M + \Om_{\Lm}} \nn \\ & = & - \Om_{\Lm} = - 0.692 \: . \eea 

\begin{figure}[ht!]
\centerline{\mbox {\epsfxsize=9.cm \epsfysize=7.cm
\epsfbox{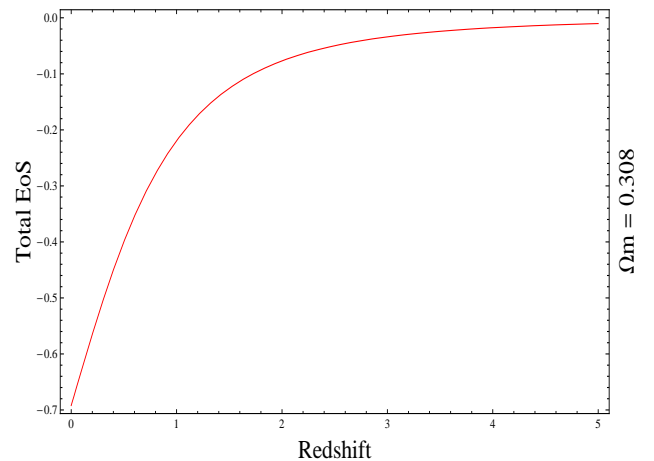}}} \caption{Total EoS parameter, $w_{tot}$, as function of the cosmological redshift, $z$, in the context of the $\Lm$CDM-like (i.e., $\Gm = 0$) limit of the polytropic DM model. We note that, today, $w_{tot} \approx - 0.7$, while, for $z \geq 3$, it remains very close to zero, as it is suggested by $\Lm$CDM cosmology.}
\end{figure} 

For $\Gm \neq 0$, several physical requirements in the context of the polytropic DM approach (such as a nonnegative velocity-of-sound square, $c_s^2$, and the large-scale-structure amendment for CDM) can lead to successive constraints on the polytropic exponent, the value of which, for $\Om_M = 0.308$, settles down to the range $- 0.103 < \Gm \leq 0$. For each and every value of $\Gm$ in this range, the theoretically-derived (in the framework of the polytropic DM model) expression of the luminosity distance, $d_L (z)$, fits to high accuracy the Hubble diagram of the 580 standard candles (cf. Fig. 6 of Kleidis \& Spyrou 2015) that constitute the Union 2.1 SNe Ia Compilation (Suzuki et al. 2012). At the same time, this model does not suffer from the "age problem" (cf. Fig. 2 of Kleidis \& Spyrou 2015) and significantly alleviates the "coincidence problem" [cf. Eqs. (106) - (108) of Kleidis \& Spyrou 2015].

Nevertheless, we need to stress that, it is not yet clear what sort of microphysics would produce the postulated polytropic behavior; hence, our model is to be seen as an effective (phenomenological) approach of an elementary physics scenario yet to be discovered (see, e.g., Gondolo \& Freese 2003; Arkani-Hamed et al. 2009; Van den Aarssen et al. 2012).

Finally, one would expect that, in a cosmological model with thermodynamical content, potential deviations from Hubble flow, i.e., "peculiar velocities" of the (non-relativistic) CDM units (e.g., the volume elements of the CDM fluid)\footnote{In a spatially-flat cosmological model, the proper distance, $D$, is defined as $D = R(t) r$, where $r$ is the comoving radial coordinate. Accordingly, the corresponding (proper) velocity is given by $v_{prop} \equiv \frac{d D}{d t} = H D + \ups_{com} R = \ups_{rec} + \ups_{pec}$, where $\ups_{rec} = H D$ is the recession velocity, driven by Hubble's law, and $\ups_{pec} = \ups_{com} R$ represents the "peculiar velocity" field, with $\ups_{com} = \frac{d r}{d t}$ being its comoving counterpart [see, e.g., Peacock 1999, Eq. (15.17), p. 463].} conform with the rms-velocity law of Statistical Physics, that is \be \langle \ups_{pec} \rangle = \sqrt{3 \frac{p}{\rh}} \ee (see, e.g., Landau \& Lifshitz 1969, p. 112), where the brackets denote average values. On this basis, the comoving counterpart of the "peculiar" velocity field, $\ups \equiv \langle \ups_{com} \rangle$, obeys the relation \be \langle \ups_{com} \rangle = \frac{1}{R} \sqrt{3 \frac{p}{\rh}} \: . \ee Eq. (13) [or, equivalently, Eq. (14)] is yet to be confirmed, for the polytropic CDM approach to be self-contained (in connection, however, see Sect. 4, discussion on Fig. 5). 

In view of all the above pros and cons, the polytropic CDM model under consideration can rise from the rank as a mighty contestant for a viable (though conventional) DE model. Still, however, there is a critical issue that this model should also confront with, that is, to demonstrate its compatibility with current observational data concerning structure formation.

\section{Perturbation equations in Fourier space}

The Newtonian (longitudinal) gauge, perticularly advocated by Mukhanov et al. (1992), has the advantage that the gauge-invariant quantities coincide to the corresponding physical ones. Consequently, in a model where the overall-energy density is due to an ideal fluid, it is reasonable to neglect any anisotropic stresses. In such a model, there remains only one metric perturbation undetermined, namely, the generalized Newtonian potential (see, e.g., Knobel 2012, p. 73). Accordingly, the cosmological spacetime metric of a (dark-) matter-dominated Universe perturbed over a spatially-flat FRW model is written in the form \be ds^2 = R^2 (\et) \left [ \left ( 1 - 2 \phi \right ) d \et^2 - \left ( 1 + 2 \phi \right ) \dl_{ij} dx^i dx^j \right ] \ee (see, e.g., Ma \& Bertschinger 1995), where $R (\et)$ is the scale factor in terms of conformal time, $\et = \int^t \frac{dt}{R(t)}$, Latin indices $(i, \: j = 1, \: 2, \: 3)$ refer to the three-dimensional spatial slices of the four-dimensional spacetime, $\dl_{ij}$ is the Kronecker symbol and $\phi$ is the scalar gravitational potential. It is worth noting that, in general, perturbations, $\dl g_{\mu \nu}$, over the metric tensor, $g_{\mu \nu}$, contain also contributions from vector and tensor fields, which, however, exhibit no instabilities. In an expanding Universe, vector perturbations decay kinematically, whereas tensor perturbations lead to gravitational waves, which do not couple with energy density and pressure inhomogeneities (see, e.g., Bardeen 1980). It is only the scalar perturbations that can lead to growing inhomogeneities, which, in turn, have an important effect on the dynamics of matter (see, e.g., Kodama \& Sasaki 1984). 

The metric perturbation, $\phi$, is a solution to the perturbed Einstein equations \be \dl {\cal G}_{\mu \nu} = - 8 \pi G \dl {\cal T}_{\mu \nu} \: , \ee where $\dl {\cal G}_{\mu \nu}$ are small-scale fluctuations of the homonymous tensor, ${\cal G}_{\mu \nu}$, and $\dl {\cal T}_{\mu \nu}$ is the energy-momentum tensor of the Universe matter-energy content perturbed over a perfect-fluid source \be {\cal T}_{\mu \nu} = (\varep + p) u_{\mu} u_{\nu} - g_{\mu \nu} p \: . \ee In Eq. (17), \be u^{\mu} = \frac{d x^{\mu}}{ds} = \left ( \frac{1}{R} , 0, 0, 0 \right ) \ee is the (unperturbed) four-velocity of the fluid comoving with Universe expansion. Accordingly, deviations from Hubble flow, i.e., "peculiar" fluctuations over $u^{\mu}$ can be cast in the form \be \dl u^{\mu} = \frac{1}{R} \left ( 0, \ups^1, \ups^2, \ups^3 \right ) \: , \ee where $\ups^i = d x^i / d \et$ are the comoving components of the small (as compared to the Universe expansion rate) "peculiar" velocity of the fluid (in terms of conformal time). In this case, \be \dl u_i = g_{ij} \frac{1}{R} \ups^j = - R^2 \dl_{ij} \frac{1}{R} \ups^j + O_2 \simeq - R \ups_i \: , \ee where the symbol $O_2$ denotes terms of second order in the (small) quantities $\ups$, $\phi$, and $\frac{\dl \rh}{\rh}$, with $\dl \rh$ being the rest-mass density perturbation. As given by Eq. (20), $\dl u_i$ is dimensionless, and the total four-velocity, $U^{\mu} = u^{\mu} + \dl u^{\mu}$, satisfies the condition \be U_{\mu} U^{\mu} = \left ( u_{\mu} + \dl u_{\mu} \right ) \left ( u^{\mu} + \dl u^{\mu} \right ) = 1 + O_2 \: , \ee to linear terms in $\ups^i$. 

In this model, the evolution of cosmological perturbations (in Fourier space) is governed by the following two sets of differential equations (see, e.g., Kodama \& Sasaki 1984; Veeraraghavan \& Stebbins 1990; Mukhanov et al. 1992; Padmanabhan 1993, p. 152; Ma \& Bertschinger 1995; Mukhanov 2005, p. 300; Knobel 2012, p. 73; and references therein).

\texttt{Field equations:}

\be -k^2 \phi - 3 {\cal H} \left ( \phi^{\prime} + {\cal H} \phi \right ) = 4 \pi G R^2 \dl \varep \: , \ee \be \phi^{\prime} + {\cal H} \phi = 4 \pi G R^2 (\varep + p) \frac{\ups}{k} \: , \ee \be \phi^{\prime \prime} + 3 {\cal H} \phi^{\prime} + \left ( 2 {\cal H}^{\prime} + {\cal H}^2 \right ) \phi = 4 \pi G R^2 \dl p \: . \ee 

\texttt{Equations of motion:}

\be \left ( \partial_{\et} + 3 {\cal H} \right ) \dl \varep + 3 {\cal H} \dl p = - (\varep + p) \left (k \ups - 3 \phi^{\prime} \right ) \: , \ee \be \left ( \partial_{\et} + 4 {\cal H} \right ) \left [ \frac{\ups}{k} (\varep + p) \right ] = \dl p + (\varep + p) \phi \: . \ee In Eqs. (22) - (26), $\dl \varep$ and $\dl p$ are small-scale perturbations to the overall-energy density and pressure, respectively, $k$ is the comoving wavenumber of the pertubation modes, $\ups = \left ( \ups_1^2 + \ups_2^2 + \ups_3^2 \right )^{1/2}$ is the comoving "peculiar" velocity of the mode denoted by $k$, and we have set ${\cal H} = \frac{R^{\prime}}{R}$ with the prime denoting differentiation with respect to $\et$, $\partial_{\et}$. 

In a FRW model, the unperturbed quantities ${\cal H}$, $\varep$ and $p$ are related by the Friedmann equations (in terms of conformal time) \be {\cal H}^2 = \frac{8 \pi G}{3} \varep R^2 \ee and \be {\cal H}^{\prime} = - \frac{4 \pi G}{3} ( \varep + 3 p ) R^2 \: , \ee along with the (unperturbed) conservation law given by Eq. (2), which reduces to \be \varep^{\prime} + 3 {\cal H} ( \varep + p ) = 0 \: . \ee The combination of Eq. (29) with particles' number conservation law, \be \rh^{\prime} + 3 {\cal H} \rh = 0 \: , \ee yields \be p^{\prime} + 3 \Gm {\cal H} p = 0 \Rightarrow p R^{3 \Gm} = constant \: , \ee verifying the fundamental polytropic relation $pV^{\Gm} = constant$, where $V = R^3$ is the comoving volume element. Eq. (31) suggests that, in a cosmological model filled with polytropic (DM) fluid the pressure is a slowly increasing (since $- 0.103 \leq \Gm \leq 0$) function of $R$, hence, of $\et$, or/and $t$, as well.

In view of Eq. (4), the perturbation quantities $\dl \varep$ and $\dl p$ are no longer independent, being related by \be \dl \varep = \dl \rh + \frac{1}{\Gm - 1} \dl p \: , \ee where, by virtue of Eq. (1), \be \frac{\dl p}{p} = \Gm \frac{\dl \rh}{\rh} \: , \ee and, therefore, \be \dl \varep = \left [ 1 + \frac{\Gm}{\Gm - 1} \left ( \frac{p}{\rh} \right ) \right ]  \dl \rh \: . \ee At this point, we need to stress that, since we are interested in structure formation, we focus attention to $\dl \rh$ (i.e., on perturbations to the rest-mass density, $\rh$). The reason is that, they are associated with concentrations of mass, in contrast to their overall-energy counterparts, $\dl \varep$, which contain also the diffusive internal (dark) energy term.

To track the evolution of cosmological perturbations in the polytropic DM model under consideration, we begin with the equations of motion (25) and (26). Accordingly, the combination of Eqs.(25), (33) and (34), yields \bea && \frac{\frac{\Gm}{\Gm - 1} \left ( \frac{p}{\rh} \right )^{\prime}}{1 + \frac{\Gm}{\Gm - 1} \left ( \frac{p}{\rh} \right )} \dl \rh + \left ( \dl \rh \right )^{\prime} + 3 {\cal H} (1 + c_s^2) \dl \rh \nn \\ && = - \rh \left ( k \ups - 3 \phi^{\prime} \right ) \: . \eea Upon consideration of Eqs. (30) and (31), we obtain \be \frac{\frac{\Gm}{\Gm - 1} \left ( \frac{p}{\rh} \right )^{\prime}}{1 + \frac{\Gm}{\Gm - 1} \left ( \frac{p}{\rh} \right )} = - 3 {\cal H} c_s^2 \: , \ee where \be c_s^2 = \frac{\Gm \left ( \frac{p}{\rh} \right )}{1 + \frac{\Gm}{\Gm - 1} \left ( \frac{p}{\rh} \right )} \ee is the (polytropic) speed of sound [cf. Eq. (85) of Kleidis \& Spyrou (2015)], so that Eq. (35) is written in the form \be \left ( \dl \rh \right )^{\prime} + 3 {\cal H} \dl \rh = - \rh \left ( k \ups - 3 \phi^{\prime} \right ) \: . \ee Finally, taking (once again) into account Eq. (30), and defining the density contrast, $\dl$, as $\dl = \frac{\dl \rh}{\rh}$, Eq. (38) results in \be \dl^{\prime} = 3 \phi^{\prime} - k \ups \: , \ee which is identical to Eq. (3.107), p. 75, of Knobel (2012). 

In the same fashion, upon consideration of Eqs. (30) and (36), the combination of Eqs. (26) and (33) yields \be \ups^{\prime} + {\cal H} \left ( 1 - 3 c_s^2 \right ) \ups = k \left ( c_s^2 \dl + \phi \right ) \: . \ee Now, differentiating Eq. (39) and combining the outcome with Eq. (40), we obtain \be \dl^{\prime \prime} + k^2 c_s^2 \dl = k {\cal H} \left ( 1 - 3 c_s^2 \right ) \ups + 3 \phi^{\prime \prime} - k^2 \phi \: . \ee Finally, we may (re)use Eq. (39), to eliminate $\ups$ from the rhs of Eq. (41), which, in this case, results in \bea \dl^{\prime \prime} + {\cal H} \left ( 1 - 3 c_s^2 \right ) \dl^{\prime} + k^2 c_s^2 \dl & = & 3 \phi^{\prime \prime} + 3 {\cal H} \left ( 1 - 3 c_s^2 \right ) \phi^{\prime} \nn \\ & - & k^2 \phi \: . \eea For $c_s^2 = 0$, Eq. (42) reduces to Eq. (3.109), p. 75, of Knobel (2012), that governs the evolution of conventional (i.e., pressureless) CDM perturbations in an expanding, spatially-flat cosmological model. 

Eq. (42) is an inhomogeneous, second order differential equation, that governs the propagation of rest-mass density perturbations in the presence of the source term \be S (\et) = 3 \left [ \phi^{\prime \prime} + {\cal H} \left ( 1 - 3 c_s^2 \right ) \phi^{\prime} - \frac{k^2}{3} \phi \right ] \: , \ee which depends on the generalized Newtonian potential, $\phi$. 

To determine the functional form of $\phi$ [and, hence, of $S(\et)$, as well], we note that the combination of Eqs. (24) and (33) yields \be \phi^{\prime \prime} + 3 {\cal H} \phi^{\prime} + \left ( 2 {\cal H}^{\prime} + {\cal H}^2 \right ) \phi = \Gm \left ( \frac{p}{\rh} \right ) \left ( 4 \pi G R^2 \dl \rh \right ) \: , \ee while the combination of Eqs. (22) and (34) results in \be - \left [ k^2 \phi + 3 {\cal H} \left ( \phi^{\prime} + {\cal H} \phi \right ) \right ] = \left [ 1 + \frac{\Gm}{\Gm - 1} \left ( \frac{p}{\rh} \right ) \right ] \left ( 4 \pi G R^2 \dl \rh  \right ) . \ee Now, Eqs. (44) and (45) can be combined with each other, to give \bea && \phi^{\prime \prime} + 3 {\cal H} \phi^{\prime} + \left ( 2 {\cal H}^{\prime} + {\cal H}^2 \right ) \phi \nn \\ && = - \frac{\Gm \left ( \frac{p}{\rh} \right )}{ 1 + \frac{\Gm}{\Gm - 1} \left ( \frac{p}{\rh} \right )} \left [ k^2 \phi + 3 {\cal H} \left ( \phi^{\prime} + {\cal H} \phi \right ) \right ] \: , \eea which, in view of Eq. (37), can be written in the more convenient form \be \phi^{\prime \prime} + 3 {\cal H} \left ( 1 + c_s^2 \right ) \phi^{\prime} + \left [ k^2 c_s^2 + 2 {\cal H}^{\prime} + \left ( 1 + 3 c_s^2 \right ) {\cal H}^2 \right ] \phi = 0 \: . \ee Eq. (47) coincides to Eq. (7.51), p. 300, of Mukhanov (2005), that drives the isentropic $(\dl {\cal S} = 0)$ evolution of metric perturbations in an expanding, spatially-flat spacetime. 

For $\Gm \neq 0$, given the appropriate initial conditions, the solution, $\left \{ \phi, \dl, \ups \right \}$, to the set of simultaneous differential equations (40), (42) and (47) fully determines the evolution of cosmological perturbations in the polytropic DM model under consideration. Now, the only equation that has not been used, i.e., Eq. (23), may serve as a constraint on the exact functional form of $\left \{ \phi, \dl, \ups \right \}$. However, an analytic solution to the system of Eqs. (40), (42) and (47) is rather hard to get, being the scope of a future work. Accordingly, in order to take a first glance to the problem of cosmological perturbations in a polytropic DM model, in what follows, we focus attention to its $\Lm$CDM-like limit, i.e., to the case where $\Gm = 0 \left ( = c_s^2 \right )$. 

\section{Reduction to the $\Lm$CDM-like limit}

For $\Gm = 0$, we have $p = constant = - \vert p_0 \vert < 0$ and $\dl p = 0$ [cf. Eqs. (1), (8) and (33)]. In this case, in order to track the evolution of cosmological perturbations, we begin with the field equations (22) - (24). In fact, for $\Gm = 0$, Eqs. (22) and (24) decouple, since, now, Eq. (24) is written in the form \be \phi^{\prime \prime} + 3 {\cal H} \phi^{\prime} + \left ( 2 {\cal H}^{\prime} + {\cal H}^2 \right ) \phi = 0 \: , \ee which, upon consideration of Eqs. (27) and (28), results in \be \phi^{\prime \prime} + 3 {\cal H} \phi^{\prime} - 8 \pi G p_0 R^2 \phi = 0 \: . \ee On the other hand, by virtue of Eq. (34), Eq. (22) reads \be \dl \rh = - \frac{1}{4 \pi G R^2} \left [ k^2 \phi + 3 {\cal H} \left ( \phi^{\prime} + {\cal H} \phi \right ) \right ] \: . \ee Clearly, in order to determine the evolution of rest-mass density perturbations, $\dl \rh$, first of all, one needs to solve Eq. (49), that drives the dynamics of $\phi$.

To do so, in Eq. (49), we change the independent variable from conformal time, $d \et$, to its cosmic counterpart, $dt = R d \et$, yielding \be \ddot{\phi} + 4 H \dot{\phi} - 8 \pi G p_0 \phi = 0\: , \ee and once again, this time from cosmic time, $t$, to cosmological redshift, $z$, as follows \be w = 1 + z = \frac{R_0}{R (t)} \: , \ee yielding, \bea \dot{\phi} & = & - w H \frac{d \phi}{d w} ~~~\mbox{and} \nn \\ \ddot{\phi} & = & w H^2 \frac{d \phi}{d w} + \frac{1}{2} w^2 \frac{d \left ( H^2 \right )}{d w} \frac{d \phi}{d w} + w^2 H^2 \frac{d^2 \phi}{d w^2} \: . \eea Now, in terms of $w$, Eq. (51) is written in the form \be \frac{d^2 \phi}{d w^2} + \left [ \frac{1}{2 H^2} \frac{d \left ( H^2 \right )}{d w} - \frac{3}{w} \right ] \frac{ d \phi}{d w} - \frac{8 \pi G p_0}{w^2 H^2} \phi = 0 \: . \ee In view of Eq. (9), in the $\Lm$CDM-like $(\Gm= 0)$ limit of the polytropic DM model under consideration, the Hubble parameter is given by \be H^2 = H_0^2 \left [ \Om_M w^3 + \left ( 1 - \Om_M \right ) \right ] \: , \ee so that, eventually, Eq. (54) results in \bea \frac{d^2 \phi}{d w^2} & + & \left [ \frac{3/2}{w} \frac{1}{\left( 1 + \frac{1 - \Om_M}{\Om_M} \frac{1}{w^3} \right )} - \frac{3}{w} \right ] \frac{d \phi}{d w} \nn \\ & - & \frac{8 \pi G p_0}{\Om_M H_0^2} \frac{1}{w^5} \frac{1}{\left( 1 + \frac{1 - \Om_M}{\Om_M} \frac{1}{w^3} \right )} \phi = 0 \: . \eea At this point, we note that, the $w$-span we are interested in solving Eq. (56) falls into the range $101 \leq w \leq 1091$, corresponding to values of cosmological redshift that range from recombination $(z \approx 1090)$ to $z = 100$, when, as it is now admitted, the DM structures had already been formed (see, e.g., Naoz \& Barkana 2005; Knobel 2012, p. 76; see also Sandvik et al. 2004, for a slightly different explanation). Admitting, once again, that $\Om_M = 0.308$ (Ade et al. 2016), for every $z \geq 100$, the quantity \be \left ( w, \Om_M \right ) = \frac{1 - \Om_M}{\Om_M} \frac{1}{w^3} \leq 2.2 \times 10^{-6} \ll 1 \: , \ee is extremely small and, therefore, it can be ignored as compared to unity\footnote{Notice that, such an approximation would be accurate enough even for $z \geq 11$, i.e., at every pre-reionization epoch, since, then, $ \left ( w, \Om_M \right ) \leq 1.3 \times 10^{-3} \ll 1$, as well.}. Accordingly, Eq. (56) is reduced to \be \frac{d^2 \phi}{d w^2} - \frac{3/2}{w} \frac{d \phi}{d w} + 3 \frac{\vert p_0 \vert}{\rh_0} \frac{1}{w^5} \phi = 0 \: , \ee where we have taken into account that, in view of Eq. (23) of Kleidis \& Spyrou (2015), we have \be - \frac{8 \pi G p_0}{\Om_M H_0^2} = 3 \frac{\vert p_0 \vert}{\rh_0} \left [ = 3 \frac{1 - \Om_M}{\Om_M} \right ] \: . \ee Eq. (58) admits the solution \be \phi (w) = w^{5/4} Z_{\frac{5}{6}} \left ( 2 \sqrt{\frac{\vert p_0 \vert}{3 \rh_0}} w^{-3/2} \right ) \ee (see, e.g., Gradshteyn \& Ryzhik 2007, Eq. 8.491.12, p. 932), where \bea Z_{\frac{5}{6}} \left ( 2 \sqrt{\frac{\vert p_0 \vert}{3 \rh_0}} w^{-3/2} \right ) & = & {\cal C}_1 (k) J_{\frac{5}{6}} \left ( 2 \sqrt{\frac{\vert p_0 \vert}{3 \rh_0}} w^{-3/2} \right ) \nn \\ & + & {\cal C}_2 (k) Y_{\frac{5}{6}} \left ( 2 \sqrt{\frac{\vert p_0 \vert}{3 \rh_0}} w^{-3/2} \right ) \eea is the linear combination of Bessel functions of the first, $J_{\nu}$, and the second, $Y_{\nu}$, kind, of order $\nu = \frac{5}{6}$, and ${\cal C}_1 (k)$, ${\cal C}_2 (k)$ are (complex) constants. On physical grounds, we admit that ${\cal C}_2 (k) = 0$, otherwise, at large $z$, we would have $\phi \rarrow \infty$, something that is not anticipated, not even by inflationary cosmology (see, e.g., Linde 1990, p. 136). In this case, at large values of $w$, the argument $(\sim w^{-3/2})$ of the only Bessel function left, $J_{\frac{5}{6}}$, diminishes, so that, \be J_{\frac{5}{6}} (\arg) \rarrow \frac{6}{5 \Gm \left ( \frac{5}{6} \right )} \left ( \frac{\arg}{2} \right )^{5/6} \ee (see, e.g., Olver et al. 2010, Eq. 10.7.3, p. 223), where $\Gm \left ( \frac{5}{6} \right ) = 1.12879$ is Euler's Gamma function of argument equal to $\frac{5}{6}$. Accordingly, \be \phi (w \gg 1) \longrightarrow \frac{6}{5} {\cal C}_1 (k) \frac{1}{\Gm \left ( \frac{5}{6} \right ) } \left ( \frac{ \vert p_0 \vert}{3 \rh_0} \right )^{5/12} = constant \: . \ee The amplitude of metric perturbations, $\vert \phi \vert$, normalized over $\vert {\cal C}_1 (k) \vert$, as a function of $w$, is given in Fig. 2. Notice that, for $z \geq 100$, we have $\vert \phi \vert \approx constant$, in complete correspondence to Eq. (63). In this way, the solution given by Eq. (60) justifies the scientific perception that, in the early matter-dominated era metric perturbations were (more or less) constant (see, e.g., Knobel 2012, p. 75).

\begin{figure}[ht!]
\centerline{\mbox {\epsfxsize=9.cm \epsfysize=7.cm
\epsfbox{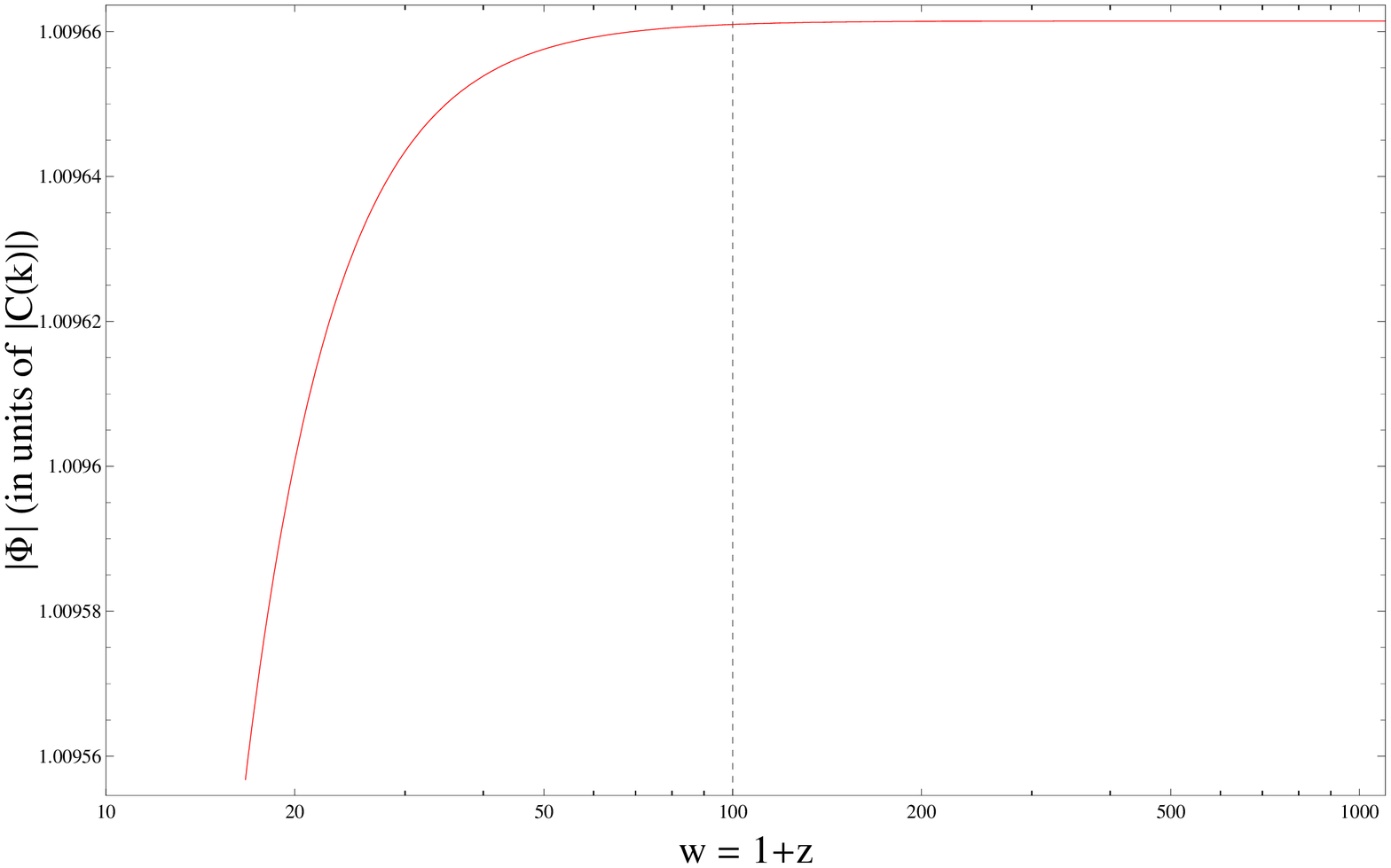}}} \caption{Amplitude of the generalized Newtonian potential, $\vert \phi \vert$, normalized over $\vert {\cal C}_1 (k) \vert$, as a function of $w = 1 + z$. It is evident that, for $w \geq 100$, $\vert \phi \vert = constant$ to high accuracy.}
\end{figure}

Now, by virtue of Eq. (60), Eq. (50) exhibits the evolution of rest-mass density perturbations in terms of $w = 1+z$, namely, \be \dl \rh = \frac{1}{4 \pi G} \left [ 3 w H^2 \frac{d \phi}{d w} - \left ( k_0^{ph} \right )^2 w^2 \phi - 3 H^2 \phi \right ] \: , \ee where we have defined the physical wavenumber at the present epoch, $k_0^{ph}$, as $k_0^{ph} = \frac{k}{R_0}$. Inserting Eq. (55) into Eq. (64) and taking into account condition (57), we obtain \be \dl \rh = 2 \rh_0 \left [ w^4 \frac{d \phi}{d w} - \frac{1}{3 \Om_M} \left ( \frac{k_0^{ph}}{H_0} \right )^2 w^2 \phi - w^3 \phi \right ] \: , \ee where, in view of Eq. (23) of Kleidis \& Spyrou (2015), we have set $8 \pi G \rh_0 = 3 \Om_M H_0^2$. Notice that, in Eq. (65), the quantity \be \frac{k_0^{ph}}{H_0} = 2 \pi \frac{\ell_{H_0}}{\lm_0^{ph}} \ee is a measure of the number of potential structures (each one of linear dimension $\lm_0^{ph}$) inside the disk of present-time (Hubble) radius $\ell_{H_0} = \frac{1}{H_0} \approx 4.28 \: Gpc$. In fact, depending on the linear dimensions of the various structures observed today, the number given by Eq. (66) may vary from the order of ten (e.g., as regards galaxy filaments) to $10^5$ (as regards structures of linear dimensions comparable to those of individual galaxies). Clearly, the smaller the linear dimensions of a particular perturbation's scale is, the more numerous its representatives at the present epoch will be. 

Eventually, dividing both parts of Eq. (65) by $\rh$ and taking into account Eq. (5), we find that the rest-mass density contrast, $\dl = \frac{\dl \rh}{\rh}$, is related to metric perturbation, $\phi$, as \be \dl = 2 \left [ w \frac{d \phi}{dw} - \frac{1}{3 \Om_M} \left ( \frac{k_0^{ph}}{H_0} \right )^2 \frac{1}{w} \phi - \phi \right ] \: . \ee Upon consideration of Eqs. (60) and (61), and taking into account Gradshteyn \& Ryzhik (2007), Eq. 8.472.2, p. 926, Eq. (67) results in \bea \dl = \frac{2}{w^{1/4}} {\cal C}_1 (k) \left [ \sqrt{\frac{3 \vert p_0 \vert}{\rh_0}} J_{\frac{11}{6}} \left ( 2 \sqrt{\frac{\vert p_0 \vert}{3 \rh_0}} w^{-3/2} \right ) \right . \nn \\- \left. \frac{1}{3 \Om_M} \left ( \frac{k_0^{ph}}{H_0} \right )^2 w^{1/2} J_{\frac{5}{6}} \left ( 2 \sqrt{\frac{\vert p_0 \vert}{3 \rh_0}} w^{-3/2} \right ) \right. \nn \\ \left. - w^{3/2} J_{\frac{5}{6}} \left ( 2 \sqrt{\frac{\vert p_0 \vert}{3 \rh_0}} w^{-3/2} \right ) \right ] \: . \eea At large $z$ $(w \gg 1)$, upon consideration of the constraint (57) and Eq. (62), we obtain \bea \dl (w \gg 1) & \longrightarrow & 2 {\cal C}_1 (k) \frac{1}{\Gm \left ( \frac{11}{6} \right ) } \left ( \frac{ \vert p_0 \vert}{3 \rh_0} \right )^{5/12} \nn \\ & \times & \left [ 1 + \frac{1}{3 \Om_M} \left ( \frac{k_0^{ph}}{H_0} \right )^2 \frac{1}{w} \right ] \: . \eea 

Now, as regards Eq. (69), there are two points worth noting: 

First, as far as small-scale structures (from the cosmological point of view) are concerned, $\left ( \frac{k_0^{ph}}{H_0} \right ) \gg 1$, i.e., the second term in brackets on the rhs of Eq. (69) is dominant for every $w$. On the contrary, for $k_0^{ph} \approx H_0$, i.e., as regards the very-large-scale structures we see today [e.g., the Hercules - Corona Borealis Great Wall, a huge filament measuring more than $10^{10}$ light years across (Horvath et al. 2013)], this term is suppressed at large values of $w$, as $w^{-1}$. Indeed, from Fig. 3 we observe that, the larger the present-time scale of a structure is, the lower the growth rate (and the amplitude) of the associated perturbation mode will be. In other words, according to the polytropic approach to a $\Lm$CDM-like model, creation of the large-scale structures we see today is disfavoured at early times (at large values of $w$), suggesting that, small-scale structures must have been formed first, aggregating to form larger structures later; that is, exactly what the CDM scenario implies. 

\begin{figure}[ht!]
\centerline{\mbox {\epsfxsize=9.cm \epsfysize=7.cm
\epsfbox{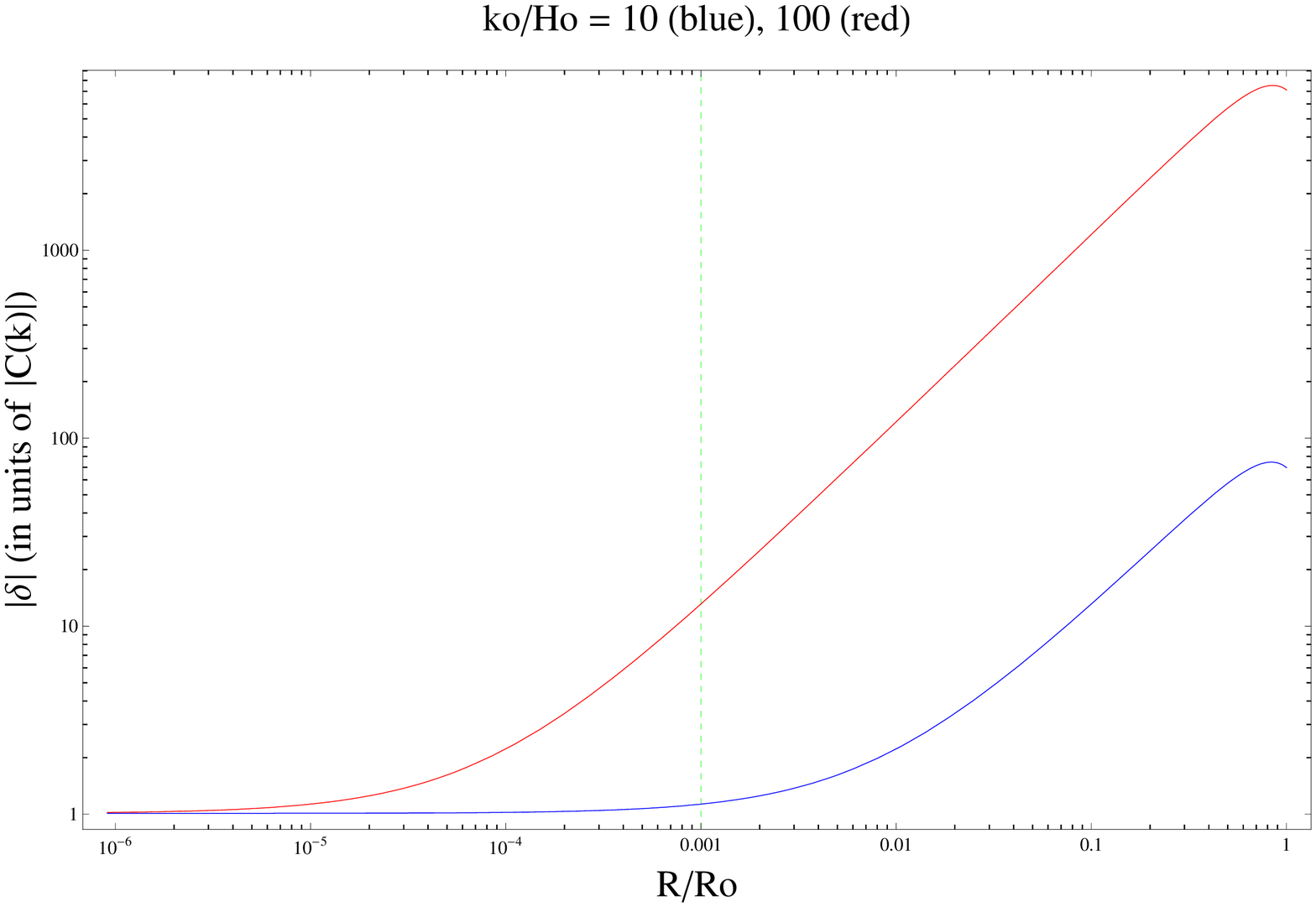}}} \caption{Amplitude of the rest-mass density contrast, $\vert \dl \vert$, normalized over $\vert {\cal C}_1 (k) \vert$, as a function of the Universe scale factor, $R$, for $\left ( \frac{k_0^{ph}}{H_0} \right ) = 10$ (blue solid line) and $100$ (red solid line). The vertical dashed line marks the recombination epoch. We observe that, the larger the present-time scale of a structure is, the more suppressed the associated perturbation mode will be.}
\end{figure}

The second point is that, at every post-recombination epoch $(z < 1090)$, to leading order in $\left ( \frac{k_0^{ph}}{H_0} \right )$ (i.e., as regards small-scale perturbation modes) Eq. (69) is reduced to \be \dl \sim w^{-1} \sim R \: . \ee In view of Eq. (70), the evolution of matter perturbations in the $\Lm$CDM-like limit of the polytropic-DM model under consideration conforms, also, with the so-called Meszaros effect (Meszaros 1974), that is admitted to govern the evolution of cosmological perturbations during the matter-dominated epoch; namely, the rest-mass density contrast grows smoothly, being proportional to the Universe scale factor, $R$ (cf. Fig. 3). Notice that, this effect applies much more accurately to small-scale structures, for which $\left ( \frac{k_0^{ph}}{H_0} \right ) \gg 1$. Indeed, for $\left ( \frac{k_0^{ph}}{H_0} \right ) = 1000$, $\vert \dl \vert \sim R^{0.995}$, while for $\left ( \frac{k_0^{ph}}{H_0} \right ) = 10$, $\vert \dl \vert \sim R^{0.945}$ (cf. Fig. 4). 

\begin{figure}[ht!]
\centerline{\mbox {\epsfxsize=9.cm \epsfysize=7.cm
\epsfbox{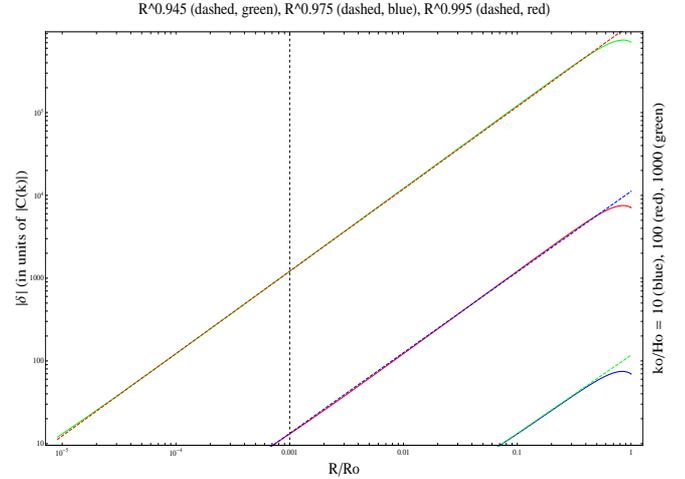}}} \caption{Amplitude of the rest-mass density contrast, $\vert \dl \vert$, in units of $\vert {\cal C}_1 (k) \vert$, as a function of the Universe scale factor, $R$, for $\left ( \frac{k_0^{ph}}{H_0} \right ) = 10$ (blue solid line), $100$ (red solid line) and $1000$ (green solid line). The vertical dashed line marks the onset of matter-dominated era, in which our (linear) petrurbative analysis holds. We observe that, for $z< 1090$, polytropic DM perturbations conform with Meszaros effect (i.e., $\dl \sim R$) to high accuracy.}
\end{figure}

Finally, by virtue of Eq. (60), we may determine, also, the comoving counterpart, $\ups$, of the "peculiar" velocity field, $\ups_{pec} = \ups R$. To do so, first, we express Eq. (23) in terms of $w$, to obtain \be \ups = \frac{2}{3} \frac{1}{\sqrt{\Om_M}} \left ( \frac{k_0^{ph}}{H_0} \right ) \frac{1}{\sqrt{w}} \left ( \phi - w \frac{d \phi}{dw} \right ) \: , \ee where, once again, we have taken into account Eq. (57). Upon consideration of Eq. (60), Eq. (71) results in \bea \ups & = & \frac{2}{3} {\cal C}_1 (k) \frac{1}{\sqrt{\Om_M}} \left ( \frac{k_0^{ph}}{H_0} \right ) \frac{1}{w^{5/4}} \left [ w^2 J_{\frac{5}{6}} \left ( 2 \sqrt{\frac{\vert p_0 \vert}{3 \rh_0}} w^{-3/2} \right ) \right. \nn \\ & - & \left. \sqrt{\frac{3 \vert p_0 \vert}{\rh_0}} J_{\frac{11}{6}} \left ( 2 \sqrt{\frac{\vert p_0 \vert}{3 \rh_0}} w^{-3/2} \right ) \right ] . \eea By virtue of Eq. (62), in the early matter-dominated era (i.e., for $w \gg 1$) the peculiar velocity (72) of the perturbation mode denoted by $k$ $(\ups \equiv \ups_k)$ is reduced to \be \ups_k \rarrow \frac{4}{5} {\cal C}_1 (k) \frac{1}{\sqrt{\Om_M}} \left ( \frac{k_0^{ph}}{H_0} \right ) \frac{1}{\Gm \left ( \frac{5}{6} \right )} \left ( \frac{\vert p_0 \vert}{3 \rh_0} \right )^{\frac{5}{12}} \frac{1}{\sqrt{w}} \: . \ee The average velocity of all perturbation modes inside the Hubble radius can be defined as \be \langle \ups_{pec} \rangle = \lim_{k \rarrow \infty} \frac{1}{k - H} \int_H^k \ups_{k^{\prime}} d k^{\prime} \: , \ee which, upon consideration of Eq. (74), yields \be \langle \ups_{pec} \rangle = \frac{4}{5 \Gm \left ( \frac{5}{6} \right )} \frac{1}{\sqrt{\Om_M}} \frac{1}{R_0 H_0} \left ( \frac{\vert p_0 \vert}{3 \rh_0} \right )^{\frac{5}{12}} \frac{1}{\sqrt{w}} \langle k {\cal C}_1 (k) \rangle \: , \ee where the (average) value $\langle k {\cal C}_1 (k) \rangle$ is determined along the lines of Eq. (74). In view of Eq. (75), because of the non-zero polytropic pressure $(p_0 \neq 0)$, the comoving "peculiar" velocities are no longer redshifted away by cosmic expansion (i.e., by Hubble drag) as it is predicted by conventional (i.e., pressureless) structure formation theory (see, e.g., Peacock 1999, p. 470; Sparke \& Ghallagher 2007, p. 350). Instead, on descending values of $w$, $\langle \ups_{pec} \rangle$ increases as \be \langle \ups_{pec} \rangle \sim \frac{1}{\sqrt{w}} \sim \left ( \frac{R}{R_0} \right )^{1/2} \ee (cf. Fig. 5). This result might be a suitable tool for either conceding or rejecting the polytropic DM model under consideration (at least, its $\Lm$CDM-like limit). In fact, for $\vert p \vert = constant = \vert p_0 \vert$ and $\rh \sim R^{-3}$, Eq. (76) is in complete correspondence to Eq. (14), regarding the evolution of comoving "peculiar" velocities in a cosmological model with thermodynamical content. This result suggests that, in the linear regime, "peculiar" velocities of the small-scale CDM concentrations conform with the (non-relativistic) rms-velocity law of conventional Statistical Physics, given by Eq. (13). It is worth noting that, although $\langle \ups_{pec} \rangle$ increases, it still remains conveniently small, e.g., for $z \geq 100$, $\langle \ups_{pec} \rangle \leq 1.3 \times 10^{-3} \ll 1$ [cf. Eq. (13)]. In fact, $\langle \ups_{pec} \rangle < 1$, at all cosmological redshifts larger than a particular value, \be z_{nlr} = \left ( 3 \frac{1 - \Om_M}{\Om_M} \right )^{1/2} - 1 \approx 0.89 \: , \ee where the suffix "nlr" stands for "non-linear regime". Indeed, $z_{nlr}$ can be considered as representing the far outer edge of the linear regime [lying safely outside the $\Lm$CDM-oriented acceleration era, which commences at $z_{tr} \approx 0.65$ (Ade et al. 2016)]. 

Notice also that, the larger the linear dimensions of a particular type of structures at the present epoch are, the smaller the "peculiar" velocity of the corresponding perturbations will be. For instance, at $z = 4$ (vertical dashed line in Fig. 5), perturbations of the order of $\left ( \frac{k_0^{ ph}}{H_0} \right ) = 1000$ (i.e., of linear dimensions comparable to those of the Coma Cluster at the present epoch) were moving at a threefold speed as compared to those of linear dimensions of the order of present-time galaxy super-clusters, i.e., of $\left ( \frac{k_0^{ph}}{H_0} \right ) = 100$ (cf. horizontal dashed lines in Fig. 5). This is a result that could be traced also observationally, e.g., by the forthcoming Euclid Mission (see, e.g., http://sci.esa.int/euclid). 

\begin{figure}[ht!]
\centerline{\mbox {\epsfxsize=9.cm \epsfysize=7.cm
\epsfbox{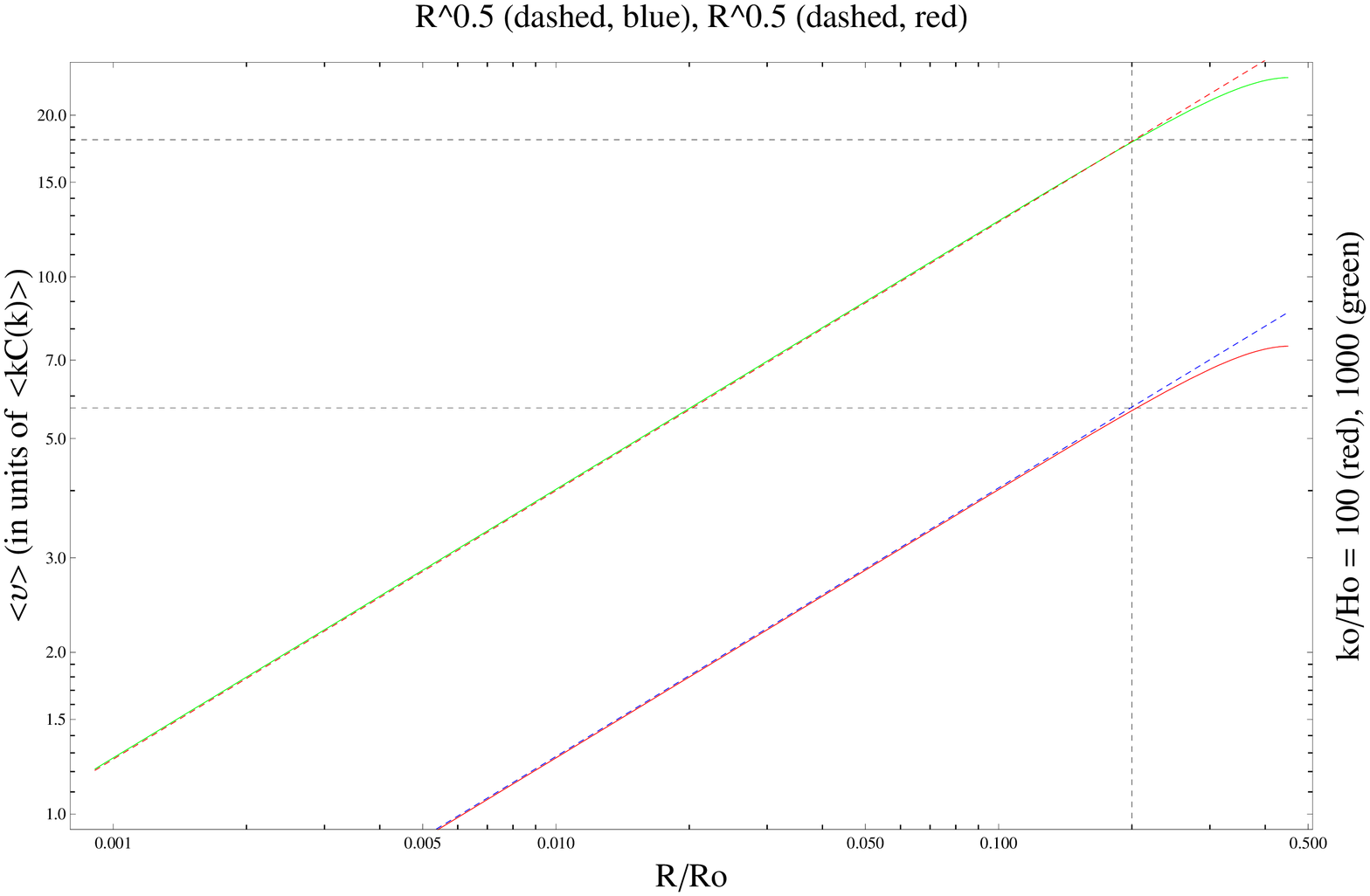}}} \caption{Average "peculiar" velocity field, $\langle \ups_{pec} \rangle$, normalized over $\langle k {\cal C}_1 (k) \rangle$, as a function of the Universe scale factor, $R$, for $\left ( \frac{k_0^{ph}}{H_0} \right ) = 100$ (red solid line) and $1000$ (green solid line). We observe that, within the linear regime, "peculiar" velocities conform with the rms-velocity law of Statistical Physics (i.e., $\langle \ups \rangle \sim \sqrt{R}$) given by Eq. (14). On both curves, the receding part arising for $\frac{R}{R_0} \geq 0.2$ $(z \leq 4)$ signals the onset of deviations from the rms-velocity law due to the collapse of linear approximation and Hubble drag.}
\end{figure}

Having used the field equations (22) - (24) to arrive at $\{ \phi, \dl, \ups \}$, the perturbed equations of motion given by Eqs. (25) and (26) may now serve as constraints to our solution. In this context, in terms of $w$, Eq. (25) is written in the form \be \frac{d \dl}{d w} = 3 \frac{d \phi}{d w} + \frac{2}{3} \frac{1}{\Om_M} \left ( \frac{k_0^{ph}}{H_0} \right )^2\frac{1}{w^2} \left ( \phi - w \frac{d \phi}{d w} \right ) \: , \ee where, once again, we have used Eq. (57). By virtue of Eq. (67), Eq. (78) results in \be 2 w \frac{d}{dw} \left ( \frac{d \phi}{dw} \right ) - 3 \frac{d \phi}{dw} = 0 \: . \ee In the same fashion, Eq. (26) yields \be \frac{1}{3} w \frac{1}{\sqrt{\Om_M}} \left ( \frac{k_0^{ph} }{H_0} \right ) \left [ 2 w \frac{d}{dw} \left ( \frac{d \phi}{dw} \right ) - 3 \frac{d \phi}{dw} \right ] = 0 \: . \ee 

\begin{figure}[ht!]
\centerline{\mbox {\epsfxsize=9.cm \epsfysize=7.cm
\epsfbox{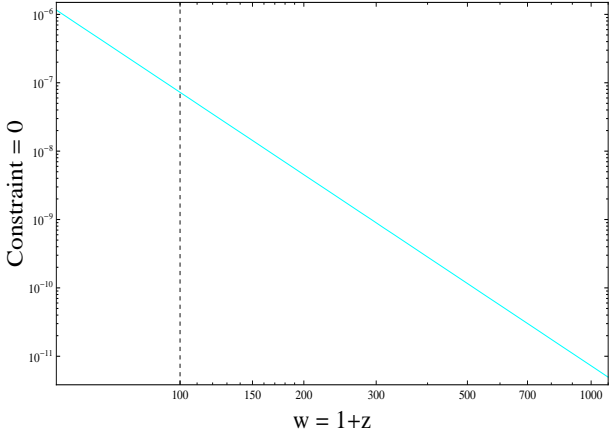}}} \caption{Equation of motion (79), being monitored over the time-span corresponding to $50 \leq w = 1+z \leq 1100$. It is evident that, for $w \geq 100$, this constraint is valid to accuracy higher than $1:10^7$.}
\end{figure}

Clearly, both constraints are reduced to one and only, namely, Eq. (79), which, for $\phi = constant$, is identically valid. Admitting that $\phi \neq constant$, we may use Eq. (60) to monitor the evolution of Eq. (79) in terms of $w$. The outcome is given in Fig. 6. We observe that, for $w \geq 100$, the constraint (79) is satisfied, to high accuracy. 

\section{The power spectrum of density perturbations}

In an isotropic cosmological model, the dimensionless power spectrum of rest-mass density perturbations is defined as \be \Dl^2(\dl) = \frac{1}{2 \pi^2} k^3 \vert \dl(k) \vert^2 \ee (see, e.g., Peacock 1999, p. 498), and, in a similar manner, the corresponding spectrum of metric perturbations is given by \be \Dl^2(\phi) = \frac{1}{2 \pi^2} k^3 \vert \phi(k) \vert^2 \ee (see, e.g., Mukhanov 2005, p. 325). Although, in principle, there is no reason why the rest-mass density spectrum should exhibit a power-law behaviour (see, e.g., Liddle \& Lyth 1993, p. 30), usually, it is admitted that \be \Dl^2 (\dl) \sim k^{3 + n_s} \: , \ee where $n_s$ is the scalar spectral index (see, e.g., Knobel 2012, p. 89). To calculate the spectrum of rest-mass density perturbations in the $\Lm$CDM-like limit of the polytropic DM model under consideration, all we need is Eq. (67). In this context, we note that \be \frac{k_0^{ph}}{H_0} = \frac{k}{R} \frac{R}{R_0} \frac{1}{H_0} = \frac{k_{ph}}{H} \frac{1}{w} \frac{H}{H_0} \: . \ee Accordingly, \bea \left ( \frac{k_0^{ph}}{H_0} \right )^2 & = & \left ( \frac{k_{ph}}{H} \right )^2 \frac{1}{w^2} \Om_M w^3 \left [ 1 + \frac{1 - \Om_M}{\Om_M} \frac {1}{w^3} \right ] \nn \\ & \simeq & \left ( \frac{k_{ph}}{H} \right )^2 \Om_M w \: , \eea where, once again, we have taken into account Eq. (57). In view of Eq. (85), Eq. (67) is written in the form \be \dl = 2 \left [ w \frac{d \phi}{dw} - \frac{1}{3} \left ( \frac{k_{ph}}{H} \right )^2 \phi - \phi \right ] \: . \ee By virtue of Eq. (86), for $\phi \approx constant$, we obtain \be \frac{\Dl^2 (\dl)}{\Dl^2 (\phi)} = 4 \left [ 1 + \frac{1}{3} \left ( \frac{k_{ph}}{H} \right )^2 \right ]^2 \: . \ee The behaviour of Eq. (87) as a function of $k_{ph}$ (measured in units of $H$) is presented in Fig. 7 (red solid line). We observe that, for $\left ( \frac{k_{ph}}{H} \right ) \geq 5$ (i.e., for every $\lm_{ph} \leq \ell_H$), the quantity $\Dl^2 (\dl) / \Dl^2 (\phi)$ exhibits a prominent power-law dependence on $k_{ph}$, of the form \be \frac{\Dl^2 (\dl)}{\Dl^2 (\phi)} = \bt \left ( \frac{k_{ph} }{H} \right )^{3.970} \: , \ee where $\bt$ is a proportionality factor. Consequently, \be \Dl^2 (\phi) \sim \frac{\Dl^2 (\dl)}{\left ( \frac{k_{ph}}{H} \right )^{3.970}} \overbrace{=}^{Eq. (81)} \frac{\left ( \frac{k_{ph}}{H} \right )^{n_s + 3}}{\left ( \frac{k_{ph}}{H} \right )^{3.970}} = \left ( \frac{k_{ph}}{H} \right )^{n_s - 0.970} . \ee 

Admitting that the power spectrum of metric perturbations is scale invariant\footnote{The requirement that $\Dl^2 (\phi) \sim k^0$ implies that, in Eqs. (61), (68) and (72), ${\cal C}_1 (k) \sim k^{-3/2}$ [cf. the combination of Eqs. (60) and (82)], in complete agreement to the (so-called) vacuum assumption, arising from inflationary cosmology amendments [cf. Liddle \& Lyth 1993, Eq. (5.34), p. 48].}, i.e., $\Dl^2 (\phi) \sim k^0$, as implied by CMB anisotropy measurements (see, e.g., Komatsu et al. 2009, 2011) and several other physical arguments (see, e.g., Mukhanov 2005, p. 345; Padmanabhan 1993, p. 229; Peacock 1999, p. 499), Eq. (89) yields \be n_s = 0.970 \: . \ee In view of Eqs. (83) and (90), we conclude that, although in principle there is no reason why the rest-mass density spectrum should exhibit a power-law behaviour, in the context of the polytropic DM model under consideration, it effectively does so, i.e., \be \Dl^2 (\dl) \sim k_{ph}^{3 + n_s^{eff}}, ~\mbox{with} ~~ n_s^{eff} = 0.970 \: . \ee Notice that the effective scalar spectral index given by Eq. (91) is a little bit lower than unity, where the value $n_s = 1$ corresponds to the Harrison-Zel'dovich-Peebles (scale invariant) spectrum, for which $\Dl^2 (\dl) \sim k^4$, i.e., $\vert \dl \vert^2 \sim k$ (Harrison 1970; Peebles \& Yu 1970; Zel'dovich \& Novikov 1970). It is now well-established that, in realistic cosmology, more power is attributed to large scales, i.e., $n_s < 1$, at $5 \sg$ confidence level (see, e.g., Bennet et al. 2013; Li et al. 2013), in complete agreement to amendments that root as back as slow-roll inflation (see, e.g., Liddle \& Lyth 2000, p. 188; Springel et al. 2006). In this context, the theoretically-derived value (91) regarding the effective scalar spectral index of rest-mass density perturbations in a polytropic DM model of constant pressure actually reproduces the corresponding observational (Planck) result, $n_s^{obs} = 0.968 \pm 0.006$ (Ade et al. 2016).

\begin{figure}[ht!]
\centerline{\mbox {\epsfxsize=9.cm \epsfysize=7.cm
\epsfbox{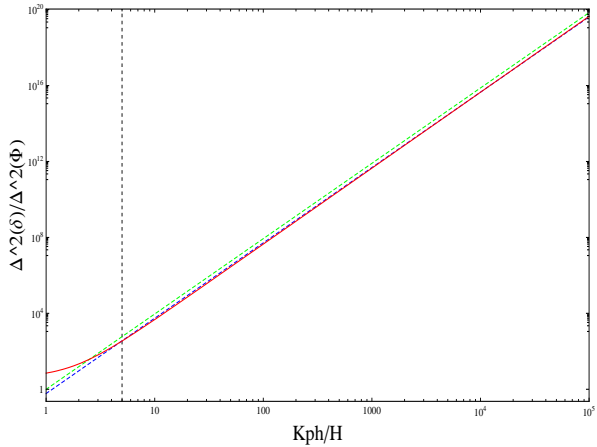}}} \caption{Plot of Eq. (87), regarding small-scale perturbations in the $\Gm = 0$ limit of a polytropic DM model (red solid line). The straight, dashed lines represent Eq. (88), for $\bt = 1$ (green dashed line) and $\bt = 0.60$ (blue dashed line), each one of slope $\al = 3.970$. It is evident that, as long as $\Dl^2 (\phi) \sim k^0$, polytropic perturbations with $\left ( \frac{k_{ph}}{H} \right ) \geq 5$ exhibit an effective power-law behaviour with scalar spectral index equal to $n_s^{eff} = 0.970$.}
\end{figure}

{\em Summarizing}, matter perturbations of linear dimensions smaller than the Hubble radius at any $t$, when accomodated in the $\Lm$CDM-like (i.e., $\Gm = 0$) limit of a polytropic DM model, effectively exhibit a power-law behaviour, of the form $\vert \dl \vert^2 \sim k^{n_s^{eff}}$, with the associated scalar spectral index being equal to $n_s^{eff} = 0.970$. To the best of our knowledge, this is the first time that a conventional model with practically zero free parameters actually predicts a theoretical result so close to observation.

\section{Discussion}

In the unified DM framework, Kleidis \& Spyrou (2015) proposed that the constituents of the cosmological dark sector (i.e., DM and DE) can (indeed) be treated as a single component when accommodated in the context of a polytropic DM fluid with thermodynamical content. In this model, macroscopically, the DE simply represents the thermodynamic energy of internal motions of the polytropic DM fluid. Depending on only one free parameter, $-0.103 < \Gm \leq 0$, the unified DM model under consideration reproduces to high accuracy the distance measurements performed with the aid of the SNe Ia standard candles, without suffering from the age problem, and significantly alleviates the coincidence problem (see, e.g., Kleidis \& Spyrou 2016). Furthermore, in its $\Gm = 0$ ($\Lm$CDM-like) limit, this model fully compromises the currently admitted behaviour of the total EoS parameter in terms of $z$ (cf. Fig. 1). In the present article we demonstrate that, in its $\Lm$CDM-like limit, the polytropic DM model under consideration is also compatible with current observational data concerning structure formation. To do so, we explore the evolution of cosmological perturbations in the $\Gm = 0$ $(p = constant \neq 0)$ limit of the aforementioned unified DM model. 

As we find, in this case, the differential equations that govern the evolution of cosmological perturbations decouple, so that, for $z \geq 100$, when the DM structures had already been formed (see, e.g., Sandvik et al. 2004; Naoz \& Barkana 2005; Knobel 2012, p. 76), they can be solved analytically. Accordingly, in the Newtonian gauge, we obtain the form of the generalized Newtonian potential, $\phi$, of the rest-mass density contrast, $\dl$, and of the comoving "peculiar" velocity field, $\ups$, as functions of the cosmological redshift, $z$. An analysis of these results, shows that our solution for $\lbrace \phi, \dl \rbrace$, reproduces every major effect already known from conventional (i.e., pressureless CDM) cosmological perturbations' theory, while the non-zero polytropic pressure drives the evolution of "peculiar" velocities along the lines of the rms-velocity law of conventional Statistical Physics.

In particular, for $z \geq 100$, the generalized Newtonian potential is $\vert \phi \vert \approx constant$ (cf. Fig. 2), justifying the current scientific perception that, in the vicinity of the matter-dominated era, metric perturbations were (more or less) constant (see, e.g., Knobel, p. 75). On the other hand, as far as matter perturbations are concerned, small-scale modes (i.e., those lying well-within the horizon) conform to high accuracy (cf. Fig. 4) with Meszaros effect (Meszaros 1974), while those of linear dimensions comparable to the present-time value of Hubble radius are suppressed (cf. Fig. 3) as $\left ( 1 + z \right )^{-1}$ [cf. Eq. (69)]. In other words, in the early matter-dominated era $(100 \leq z \leq 1090)$, only the small-scale structures we see today were allowed to be formed, which must have subsequently aggregated to form larger structures later (at lower values of cosmological redshift), in compatibility to the CDM approach (Bond \& Szalay 1983). Finally, for $z \geq 11$, the non-zero polytropic pressure results in the modification of the functional dependence of $\langle \ups_{pec} \rangle$ over $R$, which, on the approach to the present epoch, is no longer redshifted away, as predicted by conventional (i.e., pressureless) structure formation theory (see, e.g., Peacock 1999, p. 470; Sparke \& Ghallagher 2007, p. 350), but, instead, increases (cf. Fig. 5) in agreement to the rms-velocity law of Statistical Physics [cf. Eqs. (13) and (14)]. This result could provide a convenient tool for either conceding or rejecting the polytropic DM model under consideration, e.g., by the forthcoming Euclid Mission (see http://sci.esa.int/euclid).

Eventually, the dimensionless power spectrum of rest-mass density perturbations in the $\Gm = 0$ ($\Lm$CDM-like) limit of the unified DM model under study is examined. In this case, provided that the corresponding spectrum of metric perturbations is scale invariant, as implied by CMB anisotropy measurements (see, e.g., Komatsu et al. 2009, 2011) and several other physical arguments (see, e.g., Padmanabhan 1993, p. 229; Peacock 1999, p. 499), the polytropic DM model with $\Gm = 0$ is equivalent to a cosmological model in which, effectively, matter perturbations' spectrum exhibits a power law dependence on the (physical) wavenumber, of the form $\vert \dl \vert^2 \sim k^{n_s^{eff}}$, with the associated scalar spectral index being equal to $n_s^{eff} = 0.970$ (cf. Fig. 7). It is worth noting that, this value only slightly differs from the corresponding observational (Planck) result (Ade et al. 2016), i.e., $n_s^{obs} = 0.968 \pm 0.006$. To the best of our knowledge, this is the first time that a conventional model with practically zero free parameters reproduces theoretically an already observed result.


\begin{acknowledgements}

The authors would like to thank the anonymous referee, for his/her critical comments and useful suggestions, that greatly improved this article's final form. Financial support by the Research Committee of the Technological Education Institute of Central Macedonia at Serres, Greece, under grant SAT/ME/141216-279/11, is gratefully acknowledged. 

\end{acknowledgements}


\end{document}